\documentclass[acmsmall,screen,nonacm]{acmart}
\usepackage{listings}
\usepackage{xcolor}
\usepackage{wrapfig}
\usepackage{tikz}
\usetikzlibrary{arrows.meta,positioning,fit,calc}
\lstdefinelanguage{ST}{keywords={PROGRAM,END_PROGRAM,FUNCTION_BLOCK,END_FUNCTION_BLOCK,VAR,VAR_INPUT,VAR_OUTPUT,VAR_GLOBAL,END_VAR,IF,THEN,ELSIF,ELSE,END_IF,FOR,TO,BY,DO,END_FOR,WHILE,END_WHILE,REPEAT,UNTIL,END_REPEAT,CASE,OF,END_CASE,TRUE,FALSE,AND,OR,XOR,NOT,MOD},keywordstyle=\bfseries,sensitive=false,morecomment=[s]{(*}{*)},morecomment=[l]{//},commentstyle=\color{gray},morestring=[b]",stringstyle=\itshape}
\setcopyright{none}
\acmDOI{}
\acmISBN{}
\acmConference{}{}{}
\makeatletter \def\@mkauthors@i{\gdef\@currentauthors{}\gdef\@currentaffiliations{}\global\setbox\mktitle@bx=\vbox{\noindent\unvbox\mktitle@bx\par\medskip {\small\noindent Jinyun Chen\textsuperscript{1}, Juntao Hu\textsuperscript{2,*}, Haochen Li\textsuperscript{1}, Wentao Song\textsuperscript{1},\par\noindent Ningbo Gu\textsuperscript{2}, Jianwei Niu\textsuperscript{1}, Qing Gao\textsuperscript{1}, Rui Wang\textsuperscript{1}\par} {\footnotesize\noindent\textsuperscript{1}Beihang University, Beijing, China\par\noindent\textsuperscript{2}Hangzhou Innovation Institute, Beihang University, Hangzhou, China\par}\medskip}} \makeatother 
\begin{document}
\title[Compilation Is Not Correctness]{Compilation Is Not Correctness:
  Scan-by-Scan Evaluation of LLM-Generated PLC Programs}
\author{Jinyun Chen} \affiliation{\institution{Beihang University}\city{Beijing}\country{China}} \email{21373257@buaa.edu.cn}
\author{Juntao Hu} \authornote{Corresponding author.} \affiliation{\institution{Hangzhou Innovation Institute, Beihang University}\city{Hangzhou}\country{China}} \email{hujuntao@buaa.edu.cn}
\author{Haochen Li} \affiliation{\institution{Beihang University}\city{Beijing}\country{China}} \email{harr1son@buaa.edu.cn}
\author{Wentao Song} \affiliation{\institution{Beihang University}\city{Beijing}\country{China}} \email{songwt@buaa.edu.cn}
\author{Ningbo Gu} \affiliation{\institution{Hangzhou Innovation Institute, Beihang University}\city{Hangzhou}\country{China}} \email{guningbo@buaa.edu.cn}
\author{Jianwei Niu} \affiliation{\institution{Beihang University}\city{Beijing}\country{China}} \email{niujianwei@buaa.edu.cn}
\author{Qing Gao} \affiliation{\institution{Beihang University}\city{Beijing}\country{China}} \email{gaoqing@buaa.edu.cn}
\author{Rui Wang} \affiliation{\institution{Beihang University}\city{Beijing}\country{China}} \email{ruiking@buaa.edu.cn}
\renewcommand{\shortauthors}{Chen et al.}
\begin{abstract}
Programmable logic controllers (PLCs) automate industrial processes through repeated scans, each reading inputs, executing control logic, and updating outputs. Engineers typically write this logic in Structured Text (ST), a high-level textual language defined in IEC 61131-3. Large language models (LLMs) can now generate ST from natural-language requirements, yet compilation success and checks on selected properties or sampled scans can leave behavioral errors undetected. To close this evaluation gap, we introduce CycleBench, a benchmark of 301 ST tasks with behavioral requirements, public interfaces, reference programs, and executable scenarios. Its runtime executes each generated program and its reference under aligned inputs, initialization, and logical time, comparing their outputs after every scan. Across eight LLMs and baseline generation methods, execution pass rate trails compilation pass rate by 11.92 percentage points; observed failures include output-update, arithmetic, and timing errors. On fixed traces of programs directly generated by five LLMs, checking outputs only at every fifth scan misses 21.66\% of the 157 failures detected by per-scan comparison; across sampling phases, the missed-failure rate ranges from 12.10\% to 28.03\%. These misses stem from transient output discrepancies, whose detection depends on the sampling phase. Building on this evaluation, we propose CycleGen, which generates multiple candidate ST programs, repairs their compilation errors with bounded compiler feedback, and selects one using runtime self-tests generated in a separate model context. CycleGen improves the execution pass rate over direct generation by 8.68 percentage points. Component ablations distinguish repair and self-test selection; candidate-count analysis shows diminishing returns, with no execution-pass-rate improvement beyond five candidates in our experiments despite 38.8\% higher token consumption at eight.
\end{abstract}




\maketitle
\raggedbottom

\section{Introduction}
\label{sec:introduction}

Programmable logic controllers (PLCs) serve as core control devices in industrial manufacturing, automating industrial processes in repeated \emph{scans}: each scan reads inputs, executes control logic, and updates outputs.
Structured Text (ST)~\cite{iec61131_3_2025}, a high-level textual language defined by IEC 61131-3, is widely used for PLC programming. 
ST's Pascal-like syntax supports control algorithms, computation, and data processing; function-block instances retain state across calls~\cite{wang2023kst}.
ST is also a target of model-based code generation: Simulink PLC Coder generates ST from Simulink models, Stateflow charts, and MATLAB functions~\cite{mathworksPLCCoder}.
With advances in large language models (LLMs), generating code from natural-language requirements has become an active research direction~\cite{chen2021codex}. 
This progress has also motivated research on ST code generation for PLCs. 
However, reliably translating control requirements into behaviorally correct ST programs remains an important challenge~\cite{koziolek2023chatgpt,pfeiffer2024universal}.

To address this challenge, existing LLM-based PLC generation methods incorporate compilation, formal verification, or runtime testing into their workflows. 
LLM4PLC~\cite{fakih2024llm4plc} and Agents4PLC~\cite{liu2024agents4plc} use formal-verification feedback, AutoPLC~\cite{yang2025autoplc} repairs code using vendor-compiler diagnostics, and SemaPLC combines specification, compilation, and runtime checks~\cite{tu2026semaplc}. 
These mechanisms provide different kinds of evidence: compilation establishes that a program builds, property checking addresses the encoded properties, and runtime testing checks behavior on the executions and scans observed. 
However, differences in task sets and acceptance criteria make their reported success rates difficult to compare directly. Moreover, agreement at selected scans can conceal incorrect intermediate outputs in stateful PLC programs. This motivates a common benchmark that evaluates generated programs on shared tasks and checks their outputs against the same behavioral requirements after every scan.

\begin{figure}[!htbp]
  \centering
  \includegraphics[width=0.94\linewidth]{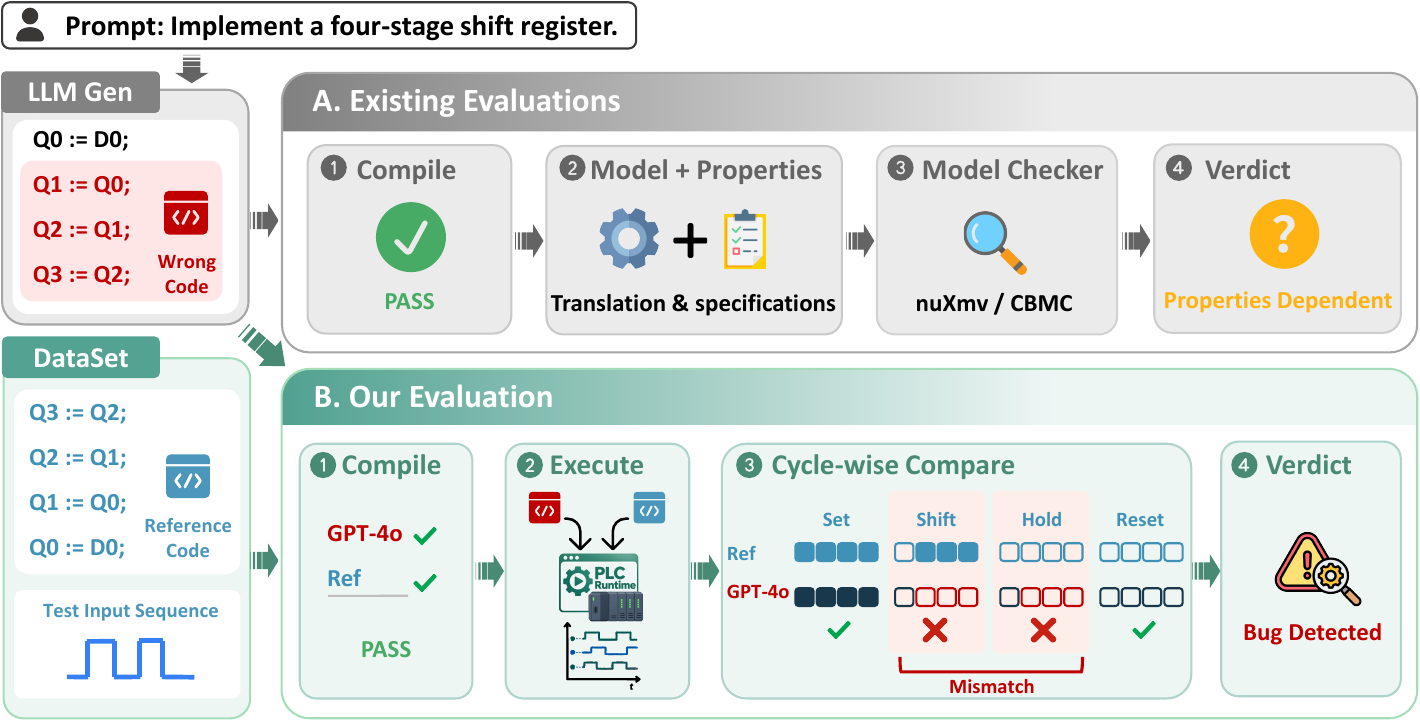}
  \caption{A shift-register failure. The reference preserves old bits;
  forward assignments propagate a new value. Panel A illustrates a
  property-checking pipeline, not a measured model-checker result. Panel B
  shows set, shift, hold, and reset; reset conceals intermediate disagreement.
  Bits are ordered \texttt{Q0}--\texttt{Q3}.}
  \Description{Reference assignments run in reverse order; generated
  assignments run forward. Panel A is a schematic property-checking
  pipeline. Panel B shows set producing 1111 in both programs, a shift with
  D0 zero yielding reference 0111 versus candidate 0000, and reset restoring
  agreement. Per-scan comparison detects the intermediate mismatch.}
  \label{fig:motivation}
\end{figure}

\label{sec:example}
Specifically, Fig.~\ref{fig:motivation} illustrates how compilation and checks on selected properties or scans can miss incorrect intermediate outputs.
On a rising edge, a shift register copies \texttt{D0} into \texttt{Q0} and the previous
\texttt{Q0}--\texttt{Q2} into \texttt{Q1}--\texttt{Q3}; set and reset override
shifting. The reference assigns outputs in reverse order to preserve old
values. A program generated directly by GPT-4o assigns them forward,
starting with \texttt{Q0 := D0}, so the new value propagates through all four
bits. After set yields \texttt{1111} at scan~11, a shift with \texttt{D0}$=0$
produces \texttt{0000} instead of \texttt{0111} at scan~12 (bits ordered
\texttt{Q0}--\texttt{Q3}); reset at scan~14 restores agreement. The program
compiles, completes execution, and ends with the reference's outputs, yet
scans~12--13 differ. A specification of the shift relation could also detect
this error: the verdict depends on which behavior is checked and when.

To measure these differences consistently, we introduce \textbf{CycleBench}: 301 ST tasks with reviewed public specifications, hidden references, and 684 scenarios spanning 61,424 scans.
A separate evaluator compares every declared output after each scan under
aligned inputs, initialization, and logical time. Across eight LLMs and
four baseline configurations, 83.99\% of 9,632 outcomes compile but only
72.07\% pass execution, a gap of 11.92 percentage points. On fixed traces
of 157 Direct behavioral failures from five models, final-scan checking
misses 98 failures; checking every fifth scan at phase zero misses 34,
even with scenario endpoints retained.
We then investigate whether runtime feedback can improve ST code generation when the generator does not have access to hidden references or evaluation scenarios.
Therefore,  we propose \textbf{CycleGen}, which adapts generated-test selection and feedback-guided repair~\cite{chen2022codet,chen2023selfdebug}.
It generates multiple candidates, repairs compilation failures, and
selects one using runtime self-tests authored beforehand in a separate
model context. The tests are withheld from code generation, and the
reference evaluator stays hidden. CycleGen raises execution pass rate
from Direct's 74.63\% to 83.31\%; controlled-pool ablations and
candidate-count analyses separate component effects and inference cost.

In particular, this work makes the following contributions:
\begin{itemize}
  \item \textbf{CycleBench.} To our knowledge, the first benchmark for
  LLM-based Structured Text generation with comprehensive per-scan
  behavioral evaluation against hidden references, comprising 301 tasks
  with reviewed specifications.
  \item \textbf{CycleGen.} A framework combining multiple candidates,
  compiler repair, and runtime self-tests, with shared-pool component
  comparisons and analyses of candidate-count gains and inference cost.
  \item \textbf{Empirical evidence.} An eight-model comparison of Direct and
  three adaptations (9,632 outcomes) quantifying the compile--execution
  gap, plus a five-model fixed-trace analysis showing how transient
  discrepancies and sampling phase hide behavioral failures.
\end{itemize}

\section{Background and Related Work}
\label{sec:background}
\label{sec:related}

\subsection{Scan Semantics and PLC Verification}
\label{sec:scan}

A cyclic PLC task invokes its program at a configured
interval~\cite{codesysTaskConfiguration}. Each scan reads inputs, executes
the control program, and updates outputs. ST statements follow the program's
control flow, so assignment order can affect later statements in the same
scan. ST organizes code into \emph{program organization units} (POUs):
functions, function blocks, and programs. Their interfaces declare inputs
(\texttt{VAR\_INPUT}), outputs (\texttt{VAR\_OUTPUT}), and read/write arguments
(\texttt{VAR\_IN\_OUT}, or inouts). Function-block instances retain internal
state across calls~\cite{iec61131_3_2025,wang2023kst}; timer blocks also depend
on elapsed time~\cite{beckhoffTOF}. Thus, behavioral testing must fix the
input sequence, initialization, time progression, and observed scans.

PLC verification provides several forms of evidence. STbmc performs
bounded temporal-property checking~\cite{lee2022stbmc}; SymPLC explores paths and task
interleavings symbolically~\cite{guo2017symplc}; K-ST gives executable ST
semantics and validates implementations differentially~\cite{wang2023kst};
and PLC simulation can step through scan cycles~\cite{blanco2020plcsim}.
These techniques analyze a given program against properties or formal
semantics. Comparing generators also requires a shared task suite with
expected behavior, which CycleBench provides.

\subsection{LLM Generation, Testing, and Repair}

General-purpose benchmarks evaluate function synthesis~\cite{chen2021codex},
library use~\cite{zhuo2024bigcodebench}, contest
problems~\cite{jain2024livecodebench}, and repository
changes~\cite{jimenez2023swebench}. EvalPlus expands test
coverage~\cite{liu2023evalplus}, and DPIaC-Eval tests infrastructure code for
deployability~\cite{zhang2026dpiac}. Reference-based execution also appears
in other code domains: VerilogEval compares simulated hardware outputs with
golden designs~\cite{liu2023verilogeval}, and SolBench uses differential
fuzzing against original smart-contract functions~\cite{chen2026solbench}.
CycleBench applies reference comparison
to PLC scan execution under controlled initialization and logical time,
and RQ4 isolates the effect of which scans are observed.

Execution feedback can also guide generation. CodeT selects among candidates
using generated tests~\cite{chen2022codet}, Self-Debugging feeds execution
results into repair~\cite{chen2023selfdebug}, and MBR-EXEC and DOCE study
execution-based selection and refinement~\cite{shi2022mbrexec,li2024doce}.
Generated expectations and self-repairs can themselves be
wrong~\cite{olausson2024selfrepair,chen2025revisit}. This motivates measuring
selection and repair separately from the reference-based final verdict.

\subsection{LLM-Based PLC Methods}

Early studies examine direct generation of PLC control
logic~\cite{koziolek2023chatgpt,pfeiffer2024universal}. Later approaches add
retrieval, specialized training, or structured control
representations~\cite{koziolek2024rag,ren2025metaindux,koziolek2025spec2control,glick2025llm4sfc}.
PLC test generation and mutation datasets provide further ways to examine
test quality~\cite{koziolek2024tests,kabir2026stmutants}.

The closest workflows combine generation with feedback. LLM4PLC uses
user-guided iteration with grammar, compiler, and SMV
checks~\cite{fakih2024llm4plc}. Agents4PLC integrates retrieval, planning,
coding, debugging, and validation agents with formal
specifications~\cite{liu2024agents4plc}. AutoPLC retrieves vendor-specific
cases and APIs and uses vendor-toolchain diagnostics for
repair~\cite{yang2025autoplc}. SemaPLC combines project or interface context
with specification, compilation, and runtime checks; its native verification
gate controls delivery~\cite{tu2026semaplc}. Its runtime strategy combines
sampled traces with persistent assertions.

\section{Benchmark: CycleBench}
\label{sec:benchmark}

Under PLC scan semantics (Section 2.1), program outputs can depend on input history, retained state, and timing. Successful compilation or agreement at selected scans can therefore leave intermediate behavioral errors undetected. To assess generated ST programs under a common behavioral criterion, we construct CycleBench. 
Its core idea is to pair reviewed public specifications with reference programs and input scenarios, and compare candidate and reference outputs after every scan under aligned inputs, public initialization, and logical time. 
CycleBench contains 301 ST function blocks and programs, with references and evaluation scenarios hidden from generation. This protocol exposes output discrepancies within the supplied scenarios.

\subsection{Benchmark Construction}
\label{sec:benchmark-construction}
\begin{figure}[!htbp]
  \centering
  \includegraphics[width=\linewidth]{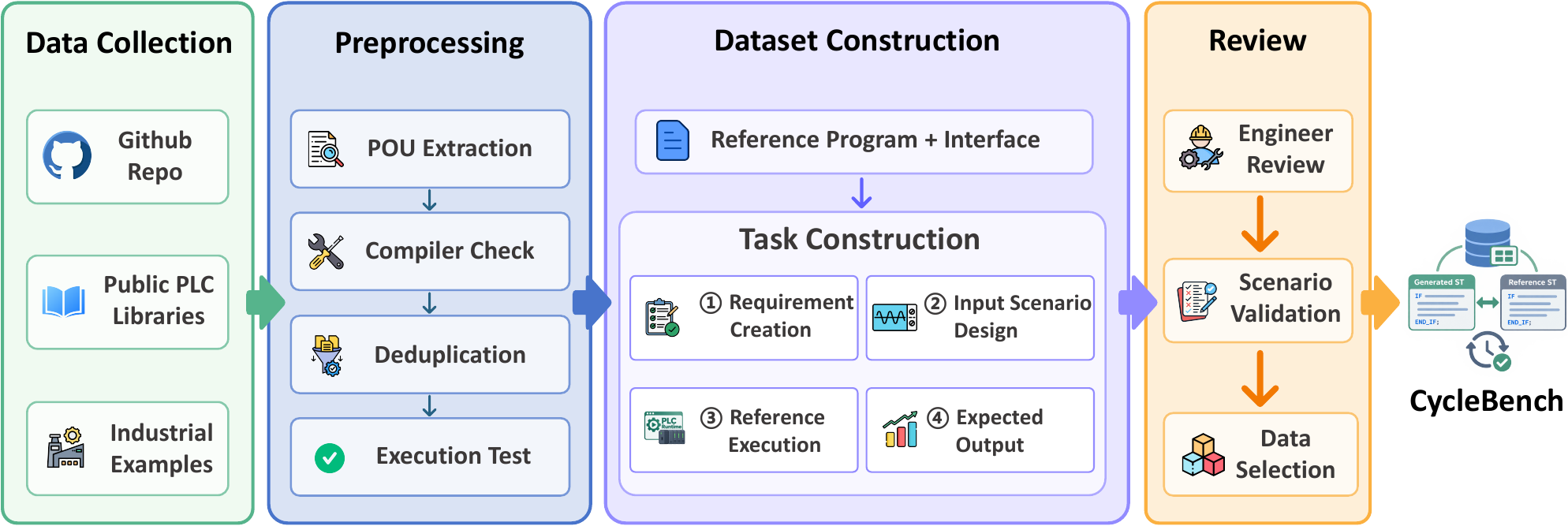}
  \caption{CycleBench construction: source collection, program preprocessing,
  task construction, and joint review by two PLC engineers with scenario
  validation and selection. Reference execution supplies the expected outputs for the
  evaluation scenarios; the final evaluation roster contains 301 tasks.}
  \label{fig:benchmark-construction}
  \Description{Four stages produce CycleBench. Sources are GitHub
  repositories (including prior PLC datasets), PLC libraries, and toolchain examples. Preprocessing
  extracts POUs, checks compilation, deduplicates code, and screens
  execution. Reference programs and interfaces support requirement
  creation and input scenario design. Reference execution supplies expected
  outputs. Joint review by two PLC engineers, scenario validation, and
  consensus-based selection determine the benchmark.}
\end{figure}

CycleBench turns existing ST implementations into reference programs and
adds a public specification and input scenarios to each.
Fig.~\ref{fig:benchmark-construction} shows the four construction stages:
data collection, preprocessing, dataset construction, and review and selection.

\noindent\textbf{\underline{Data collection.}}
We collect ST through six routes
(Table~\ref{tab:benchmark-characteristics}(b)): two GitHub collections,
AutoPLC's library, the Agents4PLC dataset, and Matiec and OpenPLC
examples~\cite{yang2025autoplc,liu2024agents4plc}.
Broad GitHub uses a wide search for ST code, whereas Curated GitHub
uses a selected set of PLC-related repositories. We recursively inspect
the repositories in both collections. The collected source files provide function blocks, programs,
and dependencies for the preprocessing stage.

\noindent\textbf{\underline{Preprocessing.}}
We extract function blocks and programs with their input, output, and
inout declarations, assemble dependencies, and check compilation with
the Matiec IEC-to-C toolchain.
Deduplication and program-family review reduce repeated
implementations~\cite{allamanis2019duplication}. Execution screening checks
whether references run under the benchmark harness, and interface checks
require observable outputs. Of 591 candidate reference programs, 12 fail
execution screening, leaving 579. Program-family deduplication and manual
selection then exclude 274, retaining 305 reference programs with
structured interfaces for dataset construction.

\noindent\textbf{\underline{Dataset construction.}}
For these 305 selected programs, we adapt existing task requirements or
draft them from reference ST with LLM assistance. They describe state transitions, priorities,
initialization, and timing; interfaces specify public signal names,
directions, and IEC types. We use
DeepSeek-V3.2 and Grok-4.6 to assist requirement drafting, and Grok-4.6
and Qwen3-Coder-Next to generate scenario inputs. Scenario
authors receive requirements, interfaces, and reference ST, but supply
no expected outputs: these come from reference execution. Scenarios
specify input sequences, scan intervals, and resets, targeting relevant
state changes, boundaries, edges, timing, and priority conflicts.

\noindent\textbf{\underline{Review and selection.}}
Two PLC engineers jointly review the behavioral requirements alongside
the reference ST, checking clarity, agreement with reference behavior,
interface completeness, and scenario suitability. Through discussion and
consensus, they refine some task prompts and the corresponding reference
code, supported by reference replay and automated checks. This review excludes four of the 305 constructed tasks because of reference-execution or public-interface contract issues. The resulting
301-task benchmark is used uniformly across all methods and models.

\noindent\textbf{\underline{Benchmark specification.}}
\label{sec:benchmark-format}
Table~\ref{tab:benchmark-format} separates the public generation
specification from the hidden evaluation suite. A method returns one
complete ST function block or program that implements the requirement
through the declared interface. The evaluator observes all declared
outputs and inout members after each scan; private
variables and internal representations may differ.

\begin{table}[htbp]
  \centering
  \caption{Content specification for CycleBench.}
  \label{tab:benchmark-format}
  \fontsize{8.5pt}{10pt}\selectfont
  \setlength{\tabcolsep}{2pt}
  \renewcommand{\arraystretch}{1.12}
  \begin{tabular}{@{}p{0.24\linewidth}p{0.23\linewidth}p{0.50\linewidth}@{}}
    \toprule
    Category & Element & Purpose \\
    \midrule
    Generation specification & Behavioral requirement & Control logic, state transitions, priorities, and timing. \\
    & POU identity & Name and kind of the function block or program. \\
    & Input/output interface & Signal names, directions, and IEC data types. \\
    & Public initial values & Starting values specified for public signals. \\
    & Functional parameters & Task-defined constants, such as thresholds or delays. \\
    \midrule
    Evaluation suite & Reference program & Executable implementation supplying expected behavior. \\
    & Input scenarios & Ordered input values for successive scans. \\
    & Execution settings & Scenario reset policy and scan interval. \\
    & Expected output traces & Reference outputs recorded after each scan. \\
    \bottomrule
  \end{tabular}
\end{table}

\subsection{Benchmark Evaluation}
\label{sec:benchmark-evaluation}

Because ST behavior unfolds over scans (Section~\ref{sec:scan}), CycleBench
compares candidate and reference outputs after every scan, exposing early,
late, or temporary discrepancies even when final outputs agree. Our runtime
translates both from ST to C with a fixed Matiec build, then compiles and
links the generated C with GCC. It drives both programs with identical
inputs, initialization, scan intervals, and reset boundaries.

\noindent\textbf{\underline{State and logical time.}}
Each scenario starts the reference and candidate in separate fresh runtime
sessions with the same public initial values; each initializes its own
private state. At every scan, the harness writes the specified inputs,
advances virtual time by one scan interval, invokes the target POU once,
and records all declared outputs and inout values. State persists between
scans; the next scenario starts afresh. Comparison includes the first scan
and does not shift traces. Logical time follows this schedule rather than
host elapsed time, so physical timing guarantees require separate
measurements. Fig.~\ref{fig:motivation} illustrates a transient disagreement.

\noindent\textbf{\underline{Task-level decision.}}
Formal task passes require preserving the public interface and POU
identity, successful compilation and execution, and matching every observed
reference value after every scan. Decoded types and values must match
exactly; arrays are rebuilt at declared bounds and compared recursively.
Surviving rounding differences fail without necessarily implying
control-logic defects; RQ2 labels them separately. All tasks remain in the
denominator. Missing candidates, interface violations, and compilation,
deployment, or runtime failures fail and are classified by stage. Evaluation
may stop at the first mismatch; passing requires complete traces for every
scenario. Each task has nonempty scenarios and outputs. RQ3's unconstrained
oracle counts behavioral matches; interface-invalid witnesses remain formal
failures.

\noindent\textbf{\underline{Isolation and validity.}}
Generation, selection, and repair cannot access references, hidden inputs,
or reference outputs. The candidate is frozen before hidden scoring.
Passing establishes agreement on the supplied finite scenarios, subject
to reference validity, rather than complete functional correctness,
termination, or physical safety~\cite{barr2015oracle,liu2023evalplus}.

\subsection{Benchmark Characteristics}
\label{sec:benchmark-characteristics}

CycleBench contains 287 function blocks and 14 programs, with 684 scenarios
and 61,424 scans (Table~\ref{tab:benchmark-characteristics}). The median task
has two inputs, two outputs, 29 reference LOC, one scenario, and 15 scans.
Repository origin is known for 259 tasks, and the three largest
repositories contribute 53.8\% of tasks. Loops, timers, edge-trigger blocks, multiway selection, and arrays
occur in 54, 32, 25, 31, and 24 tasks, respectively; these overlapping syntax
features do not establish semantic coverage. Most tasks have short
scenarios; six exceed 1,000 scans in total.

\begin{table}[!htbp]
  \centering
  \caption{Benchmark characteristics and collection routes for 301 evaluated tasks.}
  \label{tab:benchmark-characteristics}
  \fontsize{8.5pt}{10pt}\selectfont
  \setlength{\tabcolsep}{2pt}
  \renewcommand{\arraystretch}{1.08}
  \begin{minipage}[t]{0.40\linewidth}
    \vspace{0pt}
    \textbf{(a) Task and suite size}\par\smallskip
    \begin{tabular*}{\linewidth}{@{\extracolsep{\fill}}lrrr@{}}
      \toprule
      Characteristic & Min. & Med. & Max. \\
      \midrule
      Input size & 0 & 2 & 20 \\
      Output size & 1 & 2 & 18 \\
      Reference LOC & 2 & 29 & 170 \\
      Scenarios/task & 1 & 1 & 5 \\
      Scans/scenario & 1 & 15 & 10,000 \\
      Scans/task & 4 & 15 & 17,606 \\
      \midrule
      Total tasks & \multicolumn{3}{r@{}}{301} \\
      Total scenarios & \multicolumn{3}{r@{}}{684} \\
      Total scans & \multicolumn{3}{r@{}}{61,424} \\
      \bottomrule
    \end{tabular*}
  \end{minipage}\hfill
  \begin{minipage}[t]{0.58\linewidth}
    \vspace{0pt}
    \textbf{(b) Source collection}\par\smallskip
    \begin{tabular*}{\linewidth}{@{\extracolsep{\fill}}lrrr@{}}
      \toprule
      Group & Tasks & LOC (mean/med.) & Mean in/out \\
      \midrule
      Broad GitHub & 118 & 35.46/26 & 2.78/2.15 \\
      Curated GitHub & 53 & 47.28/38 & 3.81/3.45 \\
      AutoPLC library & 33 & 24.52/22 & 4.58/2.97 \\
      Agents4PLC & 90 & 39.83/34 & 2.87/2.96 \\
      Matiec examples & 4 & 22.25/22.5 & 5.25/1.75 \\
      OpenPLC examples & 3 & 30.67/25 & 3.33/1.00 \\
      \bottomrule
    \end{tabular*}
    \par\vspace{2pt}{\fontsize{8pt}{9pt}\selectfont LOC excludes blank and comment lines. Inout counts on both interface sides; an array counts as one variable. Scan totals may span resets. AutoPLC contributes 36 tasks across routes, including 33 in its library group.}
  \end{minipage}
\end{table}

\subsection{Comparison with the Agents4PLC Benchmark}
\label{sec:benchmark-comparison}

Agents4PLC's ST benchmark evaluates generated code against formal
properties~\cite{liu2024agents4plc}, rather than comparing all declared
outputs after every scan. CycleBench includes 90 tasks derived from
Agents4PLC, alongside other sources, under one per-scan evaluator. SemaPLC combines
sampled runtime traces and persistent assertions; Section~\ref{sec:design}
describes our interface-adapted comparison.

\subsection{Formative Study}
\label{sec:baseline-characterization}

We first examine the quality of LLM-generated PLC Structured Text under scan-by-scan execution.
Using the eight-model Direct results of
RQ1 (settings in Section~\ref{sec:design}), Table~\ref{tab:rq1-strata} groups
the 301 tasks by source collection and reference LOC. Pooled execution pass rate decreases from 87.17\% in the shortest
group to 48.61\% in the longest, while the compile--execution gap grows
from 3.67 to 24.07 points. All eight models perform worse in the longest
group than in the shortest, although intermediate rates need not decrease
monotonically. Every source group has a positive pooled gap, although the Matiec and
OpenPLC groups contain only four and three tasks. Execution failures thus
remain in every group, motivating CycleGen's use of behavioral evidence
during candidate selection (Section~\ref{sec:framework}).
\begin{table}[H]
  \centering
  \caption{Direct characterization by source collection (a) and reference
  length (b). Bars show each model's execution pass rate; pooled rates appear below panel (a).}
  \label{tab:rq1-strata}
  \label{tab:direct-model-loc}
  \begingroup
  \fontsize{8.5}{10.2}\selectfont
  \setlength{\tabcolsep}{1.5pt}
  \renewcommand{\arraystretch}{1.10}
  \begin{minipage}[t]{0.43\linewidth}
    \vspace{0pt}\centering
    \textbf{(a) Source collection}\par\smallskip
    \begin{tabular*}{\linewidth}{@{\extracolsep{\fill}}lrrrr@{}}
      \toprule
      Group & $n$ & Comp. & Exec. & Gap \\
      \midrule
      Broad GitHub & 118 & 87.39 & 79.77 & 7.63 \\
      Curated GitHub & 53 & 74.76 & 63.21 & 11.56 \\
      AutoPLC library & 33 & 94.32 & 90.53 & 3.79 \\
      Agents4PLC & 90 & 84.58 & 71.11 & 13.47 \\
      Matiec examples & 4 & 78.12 & 43.75 & 34.38 \\
      OpenPLC examples & 3 & 83.33 & 45.83 & 37.50 \\
      \bottomrule
    \end{tabular*}
    \par\medskip
    \textbf{Pooled rates by reference length}\par\smallskip
    \begin{tabular*}{\linewidth}{@{\extracolsep{\fill}}lrrrr@{}}
      \toprule
      LOC & 1--20 & 21--40 & 41--80 & $>$80 \\
      \midrule
      Exec. & 87.17 & 80.43 & 54.63 & 48.61 \\
      Comp. & 90.83 & 88.62 & 72.92 & 72.69 \\
      Gap & 3.67 & 8.19 & 18.29 & 24.07 \\
      \bottomrule
    \end{tabular*}
  \end{minipage}\hfill
  \begin{minipage}[t]{0.55\linewidth}
    \vspace{0pt}\centering
    \textbf{(b) Reference length}\par\smallskip
    \includegraphics[width=\linewidth]{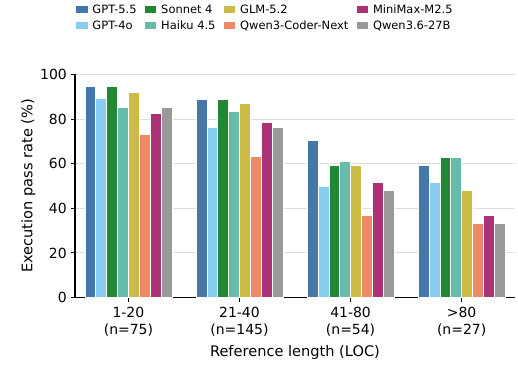}
  \end{minipage}
  \endgroup
  \par\smallskip
  \begin{minipage}{\linewidth}\footnotesize
  Comp. and Exec. are rates (\%); Gap = Comp. $-$ Exec. (pp), computed
  before rounding. $n$ is the number of tasks. Each partition covers all
  301 tasks; pooled rows contain eight model outcomes per task. Panel (b) uses the same four groups (75/145/54/27 tasks per model).
  LOC groups describe reference size, not model-derived difficulty.
  \end{minipage}
\end{table}

\section{Generation Framework: CycleGen}
\label{sec:framework}

\begin{figure}[!htbp]
  \centering
  \includegraphics[width=\linewidth]{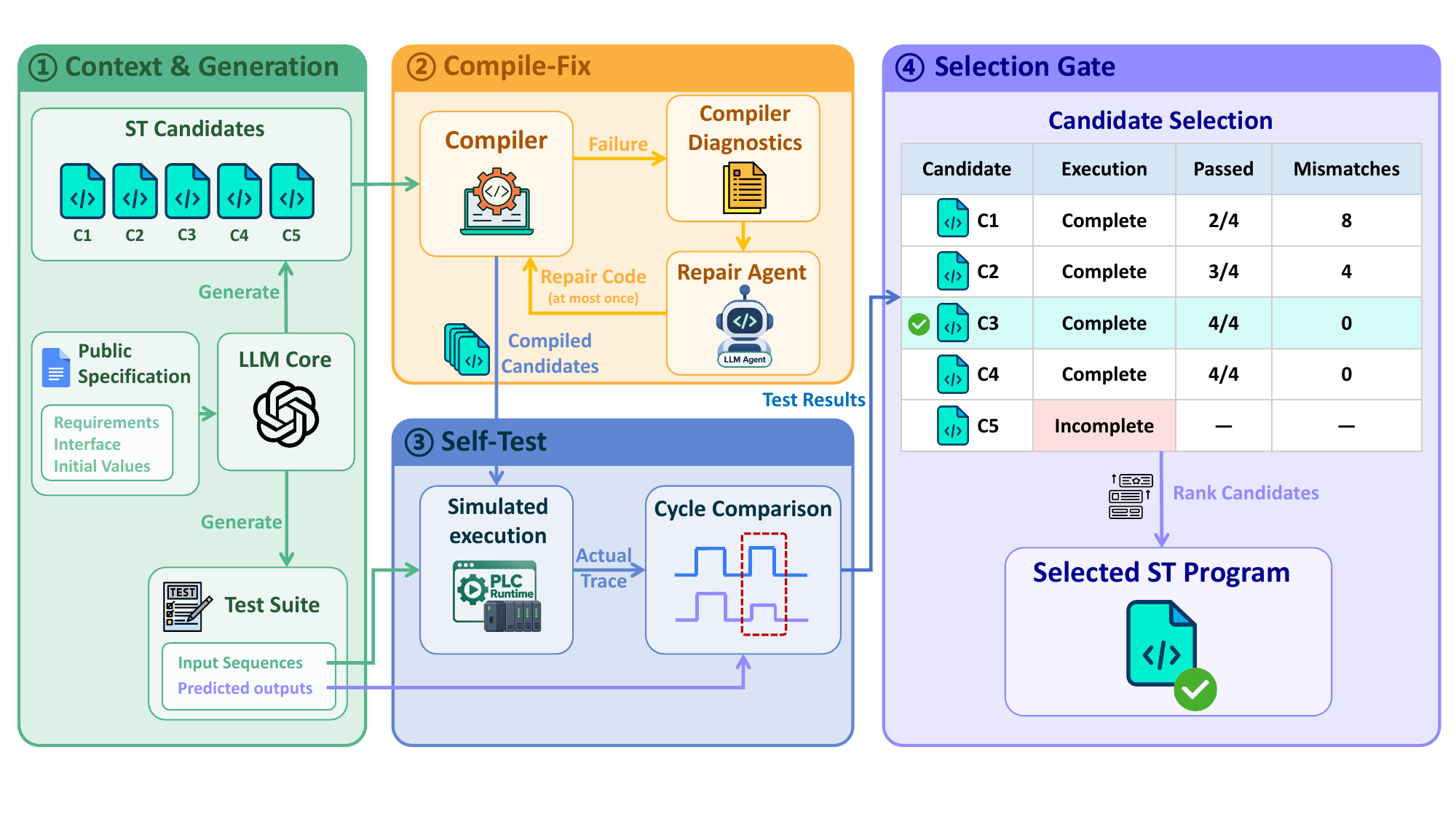}
  \caption{CycleGen's four modules use public specifications, compiler
  diagnostics, and model-generated tests to select one program. Selection
  ranks candidates without requiring all self-tests to pass. Reference-based scoring is separate from generation and selection.}
  \Description{The public specification supplies separate model contexts
  for a fixed test suite and five code samples. Compile-Fix allows one
  repair per compilation failure. Self-Test executes the shared inputs and
  compares actual with predicted outputs. Selection ranks candidates by
  execution completion, passing cases, mismatches, and generation order.}
  \label{fig:framework-overview}
\end{figure}

The formative study (Section~\ref{sec:baseline-characterization}) shows that compilable ST programs can still produce incorrect outputs. 
Runtime observations can provide additional evidence for candidate selection, but reference programs and evaluation scenarios must remain hidden during generation. The core idea of CycleGen is to supplement bounded compiler-guided repair with behavioral comparison of multiple candidates using one shared suite of model-generated scan-level tests~\cite{chen2022codet,chen2023selfdebug,zhang2026dpiac}. 
The suite is fixed before code generation and provides a fallible selection signal; the selected program is frozen before hidden evaluation. 
Fig.~\ref{fig:framework-overview} shows four modules implementing this design: Context \& Generation, Compile-Fix, Self-Test, and Selection Gate.

\subsection{Candidate and Test Generation}
\label{sec:framework-generation}

CycleGen receives the behavioral requirement, public interface, initial
values, and ST dialect rules. The same model authors tests and code in
separate contexts. Tests are fixed first and withheld from code generation;
both stages precede candidate execution.

This ordering makes the suite candidate-blind: test authorship cannot
respond to an observed implementation or its execution results. Withholding
the suite also keeps candidate generation from directly targeting its
assertions. However, both contexts use the same model and specification,
so a shared misunderstanding can still appear in code and predicted
outputs. Candidate-blind authorship therefore controls information flow,
not semantic independence. Reusing one unchanged suite makes candidates
comparable on the same input histories and assertions; it does not turn
model predictions into a correctness oracle. RQ3 measures the consequences
of this distinction through controlled selection contrasts and a separate
self-test diagnostic.

\noindent\textbf{\underline{Test generation.}}
Each case specifies a reset, an input sequence, predicted scan outputs,
and a \emph{requirement anchor}: a model-authored quotation of, or reference
to, the public requirement clause that the case is intended to test.
For example, a retention test may cite a clause requiring an output to
remain active after an input is withdrawn. Its scan sequence then activates
and withdraws that input. This annotation describes the intended behavior;
it is distinct from the input/output declarations.
Every scored scan specifies all ordinary inputs and expected observable
outputs, including inout members. An explicit inout write overwrites
its current value; an omitted write leaves the current value unchanged. One suite serves the entire pool.

Before execution, validation checks the test format, declared input/output
names and types, budgets, and consistent expectations for identical input
histories. Well-formed cases with an anchor contribute assertions to the
selection score; cases without one contribute only execution-completion
evidence. The anchor check tests whether this textual link is present,
not whether the cited clause justifies the predicted outputs. Interface
validation is a separate structural check. Suites must satisfy the extent
and variation rules in Section~\ref{sec:design}; at most one
validation-guided revision is allowed before code generation.

\noindent\textbf{\underline{Candidate generation.}}
The model generates $k=5$ complete ST programs using distinct seeds
(Section~\ref{sec:design}). Empty responses and compilation failures
consume their sampling attempt. Generation order supplies the final
tie-breaker during selection.

\subsection{Compiler-Guided Repair}
\label{sec:framework-repair}

Compile-Fix uses Matiec's \texttt{iec2c} front end to translate ST to C.
Each nonempty candidate that fails this check receives at most one repair
request containing the public specification, its code, and front-end
diagnostics. A revised POU replaces the original only if it preserves the
exact public interface and passes the same check; otherwise the original
candidate and failure status are kept. A successful repair keeps its
sampling slot. A candidate is \emph{eligible} when it preserves the
interface and passes ST-to-C translation. Interface preservation is checked
separately from compilation and requires the specified POU name and public
input/output declarations. This gate does not perform native GCC compilation
or linking. All eligible candidates use the same fixed tests.

\subsection{Runtime Self-Testing}
\label{sec:framework-selftest}

Front-end acceptance does not establish executability or correct behavior.
Self-Test attempts to build each eligible program with GCC and run the
shared test inputs in the benchmark runtime
(Section~\ref{sec:benchmark-evaluation}). Build failures yield incomplete
self-test execution. Each case starts fresh, retaining
state between scans; inputs precede each scan and logical time advances by
its interval. Self-test uses Python value equality with boolean/numeric
normalization; the final evaluator uses recursive type-and-value equality.
A scored case passes only on complete matching execution. Runtime-valid
means every scored and unscored case completes, not that predictions
are correct. Only scored cases contribute pass and mismatched output-field
counts; infrastructure faults and incomplete traces have separate statuses.
Wrong predictions can penalize correct or favor incorrect
code~\cite{koziolek2024tests,chen2025revisit}; RQ3 checks this signal against
the hidden oracle.

\subsection{Candidate Selection}
\label{sec:framework-selection}

Following execution-based selection~\cite{shi2022mbrexec,li2024doce},
CycleGen ranks eligible programs lexicographically: complete execution
first, more passing scored cases next, then fewer mismatched output-field
assertions. Exact ties are resolved by generation order. In the figure's example,
C3 wins its tie with C4 because it was generated earlier.

\noindent\textbf{\underline{Fallback and termination.}}
With no usable scored suite, selection returns the earliest eligible
program. With no eligible program after repair, it returns the first
original candidate and its observed status, including an empty response.
If every eligible program has incomplete self-test execution, the remaining
ranking criteria still select one while preserving that status.

Excluding transport retries, the budget is one test-authoring call, at
most one suite revision, $k$ code generations, and at most $k$ compiler
repairs.

\section{Experimental Settings}
\label{sec:design}
\label{sec:rqs}

To evaluate CycleBench and CycleGen, we study four research questions:
\begin{itemize}
  \setlength{\itemsep}{2pt}
  \setlength{\parsep}{0pt}
  \item \textbf{RQ1:} \emph{How Effective Is CycleGen at Generating Executable ST Programs?}
  \item \textbf{RQ2:} \emph{Which Failures Occur During Compilation and Execution?}
  \item \textbf{RQ3:} \emph{How Do CycleGen's Modules Contribute to Execution Correctness?}
  \item \textbf{RQ4:} \emph{How Much Does Observation Granularity Change Correctness Verdicts?}
\end{itemize}

\label{sec:models}
We evaluate the 301-task roster on eight model deployments accessed in
September 2026: GPT-5.5, GPT-4o, Claude Sonnet~4, Claude Haiku~4.5, and
GLM-5.2 through OpenRouter; Qwen3-Coder-Next and MiniMax-M2.5 through
Bailian Beijing; and Qwen3.6-27B through a local compatible
API~\cite{openai2026gpt55,openai2024gpt4o,anthropic2025claude4,anthropic2025haiku45,zai2026glm52}.
Provider-routed aliases identify deployments rather than immutable checkpoints.
Throughout the results, baseline names denote the adapted configurations
in Table~\ref{tab:baseline-adaptations}. Each returns one program per task/model; unsuccessful
attempts remain in the denominator. CycleGen uses pretrained models at
inference time without task-example retrieval or benchmark fine-tuning; the
full roster serves as the evaluation set.

OpenRouter upstream providers were unpinned, with default reasoning mode;
Coder-Next is non-thinking, MiniMax uses mandatory thinking, and local
Qwen disables it. RQ1/RQ3 use all eight deployments; build failures and
fixed-budget timeouts score zero.

All methods use the same 301 tasks and public interfaces, including inout
members. Direct generation and CycleGen's self-tests support inout state
updates; the three baseline adaptations receive the same declarations.

Direct uses the public requirement and interface. Table~\ref{tab:baseline-adaptations}
specifies three configurations from pinned public material:
Agents4PLC's paper/demo agent sequence~\cite{liu2024agents4plc},
LLM4PLC's prompts and grammar checker~\cite{fakih2024llm4plc}, and
AutoPLC's framework and AutoDebugger~\cite{yang2025autoplc}.
They retain compiler-feedback repair but omit formal-property checking:
the inspected releases lack a validated, task-aligned property set, and
LLM4PLC lacks end-to-end model-checker orchestration. AutoPLC's native
service checks builds and deployment. Agents4PLC does not release its
retrieval implementation and private vector store, so we reconstruct its
retrieval workflow with TF--IDF over public material. Retrieval corpora are
fixed, exclude reference-overlapping and aliasing documents before indexing,
and are checked through recorded retrievals. AutoLearn is disabled.
Results characterize these adaptations, not complete native reproductions.

\par\smallskip
\noindent\textbf{\underline{SemaPLC reconstruction.}}
We run SemaPLC's released agent core and native verifier through an
interface adapter on all 301 tasks with GPT-4o~\cite{tu2026semaplc}.
We retain trajectory checkpoints allowing at most $r=0,1,2$
failure-driven repairs per native check. The common evaluator scores
the stopped candidate regardless of native acceptance; RQ1 reports its
single-model results. Each task is bounded by 12 provider
calls, 180,000 tokens, US\$2, and 1,200 seconds. Generation stops at
native acceptance or the first applicable budget or repair limit; the
resulting candidate is frozen before common-evaluator scoring.

\begin{table}[htbp]
  \centering
  \caption{Baseline configurations from pinned public material. Listed
  methods use a 16,384-token output cap per request; budgets differ.}
  \label{tab:baseline-adaptations}
  \small
  \setlength{\tabcolsep}{3pt}
  \begin{tabular}{@{}p{0.14\linewidth}p{0.57\linewidth}p{0.23\linewidth}@{}}
    \toprule
    Method & Retained structure and substitutions & Budget \\
    \midrule
    Agents4PLC & Reconstructed retrieval, planning, coding, and compiler-feedback
    debugging; private vector store replaced by public TF--IDF;
    PLCverif/nuXmv not run. &
    At most one debug repair. \\
    LLM4PLC & Official scan/FSM planning and code prompts with compiler-feedback
    repair; SMV/plant-model loop, LoRA adapters, and interactive engineer
    loop not reconstructed. &
    One plan, at most one clarification and three code attempts. \\
    AutoPLC & Released framework with native AutoDebugger; CODESYS service
    redirected to Matiec; private online KB replaced by local BM25;
    AutoLearn disabled. &
    At most three compiler-check/debug iterations. \\
    \bottomrule
  \end{tabular}
\end{table}

Table~\ref{tab:baseline-adaptations} lists configured limits, not observed
repair counts; LLM4PLC's attempt limit is our study setting.

Initial code generation uses temperature~0.7 for Direct, CycleGen, and the
three adaptations; planning, test authoring, and repair use~0 except
AutoPLC's code editor (0.7). The adaptations use top-p~1 and seed~42;
five-candidate calls use seeds~42--46, provider-default top-p, and share
candidate one with Direct. Candidates~6--8 use seeds~47--49, respectively,
with temperature~0.7, top-p~1.0, and a 16,384-token output cap, while
preserving each model's original deployment route.
Ablations share initial candidates and candidate-blind tests; invalid suites invoke fallback
(Section~\ref{sec:framework-selection}). These settings do not isolate
temperature effects.

Model-generated self-test suites have at most six cases, ten scans per case, and forty scans total.
Admission requires at least two scored cases, four scored scans, complete
ordinary inputs and expected outputs/inouts, and two distinct written
values per writable input; omitted inout writes retain state. These fixed
filters bound cost and exclude degenerate suites.

Compilation rate (Comp.) is the fraction of task–model pairs whose returned
program completes the common evaluator’s ST-to-C translation and native GCC compilation and linking.
It is distinct from CycleGen's front-end eligibility check.
Execution pass rate is the fraction of tasks whose returned program
preserves the public interface, compiles and completes all scenarios, and
matches all declared outputs/inouts at every scan;
the gap subtracts it from compilation rate. Post-compilation deployment
and runtime failures remain distinct from behavioral mismatches.
Each condition scores one returned program, unlike pass@$k$~\cite{chen2021codex};
five candidate slots are not independent workflow runs, so this matrix
cannot estimate reliability@$k$ or seed variance~\cite{jiang2026reliability}.

RQ2 classifies the first compiler diagnostic per Direct compilation failure.
One reviewer used prompts, reference/candidate code, and mismatch evidence
to assign one primary category to each of 157 Direct behavioral failures
(69 tasks, 156 distinct programs, with one program shared by two outcomes);
numerical disagreements are separate.
To assess labeling consistency, a second ST-experienced annotator,
blinded to the original labels, independently classified one outcome from
each of 40 randomly sampled failing tasks using the same codebook. RQ2 reports agreement and unweighted Cohen's $\kappa$.

RQ2 and RQ4 use GPT-5.5, GPT-4o, Claude Sonnet~4, Claude Haiku~4.5,
and GLM-5.2, yielding 6,020 baseline outcomes and 1,505 Direct outcomes. RQ4 holds programs and traces fixed, comparing every
scan against strides two, five, and ten (including scenario endpoints), and
final-scan-only observation. All 157 failure traces have validated
observations. Missed failures and apparent pass rates isolate observation
granularity.

Beyond benchmark drafting, AI coding assistants supported script
development, analysis, and manuscript editing. Task admission and RQ2
failure classification were performed by human reviewers.

\label{sec:stats}
Overall results weight models equally. Primary RQ1/RQ3 intervals
use 20,000 paired task-cluster bootstrap samples, seed~20260911, retaining
all eight model outcomes within each resampled task~\cite{koehn2004bootstrap};
they are pointwise rather than multiplicity-adjusted simultaneous bounds.
For seven CycleGen contrasts and the unrepaired-pool selector contrast,
we sum paired pass differences across models within each task and compute
two-sided exact sign-flip $p$-values over nonzero task sums, with Holm
correction across eight tests. The 20,000 bootstrap draws estimate confidence
intervals, not these $p$-values. The tests assume independent task
clusters and sign exchangeability under the null; related tasks may violate
independence. These analyses are exploratory estimates on the evaluated
task set. For RQ4, we use 20,000 task-cluster resamples,
seed~20260927, retaining the five-model diagnostic clusters.

We count the model calls and input-plus-output tokens used by Direct and
CycleGen across all eight models, including test authoring and repairs.
Shared generations are counted once. RQ3 reports mean calls and tokens
per task--model pair; local compilation and execution are excluded.

\newcommand{\rqanswer}[2]{%
  \par\addvspace{4pt}\noindent
  \begingroup
  \setlength{\fboxsep}{3pt}%
  \setlength{\fboxrule}{0.9pt}%
  \fcolorbox{black!70}{black!4}{%
    \begin{minipage}{\dimexpr\linewidth-2\fboxsep-2\fboxrule\relax}
    \textbf{Answer to RQ#1.} \textit{#2}
    \end{minipage}}%
  \endgroup\par\addvspace{2pt}%
}

\section{Evaluation}
\label{sec:evaluation}
\suppressfloats[t]

We answer the four research questions on the 301-task roster
using the common evaluator of Section~\ref{sec:benchmark-evaluation};
model scopes follow Section~\ref{sec:design}.

\subsection{RQ1: How Effective Is CycleGen at Generating Executable ST Programs?}
\label{sec:rq1}

Table~\ref{tab:study-a} reports compilation and execution pass rates for each
model and method. Across 9,632 baseline outcomes, 83.99\% yield compilable
programs and 72.07\% pass the complete execution criterion, a difference
of 11.92 percentage points (95\% task-cluster bootstrap interval
$[9.63,14.43]$). The gap appears in every model--method
combination, including combinations with compilation rates above 99\%.

\begin{table}[!t]
  \centering
  \caption{Compilation, execution pass rate, and their gap across eight models.}
  \label{tab:study-a}
  \begingroup
  \fontsize{8.5}{10.3}\selectfont
  \renewcommand{\arraystretch}{1.17}
  \setlength{\tabcolsep}{2.0pt}
  \scalebox{0.9}{%
  \begin{tabular}{@{}l*{5}{rrr}@{}}
    \toprule
    & \multicolumn{3}{c}{\textbf{Direct}}
    & \multicolumn{3}{c}{\textbf{Agents4PLC}~\cite{liu2024agents4plc}}
    & \multicolumn{3}{c}{\textbf{LLM4PLC}~\cite{fakih2024llm4plc}}
    & \multicolumn{3}{c}{\textbf{AutoPLC}~\cite{yang2025autoplc}}
    & \multicolumn{3}{c}{\textbf{CycleGen} (Ours)} \\
    \cmidrule(lr){2-4}\cmidrule(lr){5-7}\cmidrule(lr){8-10}\cmidrule(lr){11-13}\cmidrule(l){14-16}
    \textbf{Model} & Comp. & Exec. & \textbf{Gap}
    & Comp. & Exec. & \textbf{Gap} & Comp. & Exec. & \textbf{Gap}
    & Comp. & Exec. & \textbf{Gap} & Comp. & Exec. & \textbf{Gap} \\
    \midrule
    GPT-5.5 & 96.68 & 84.39 & 12.29 & 97.67 & 85.05 & 12.62 & 99.34 & 84.05 & 15.28 & 97.34 & 83.72 & 13.62 & 99.00 & \textbf{86.71} & 12.29 \\
    GPT-4o & 84.39 & 72.76 & 11.63 & 88.70 & 75.75 & 12.96 & 89.37 & 70.43 & 18.94 & 72.43 & 60.80 & 11.63 & 93.69 & \textbf{82.72} & 10.96 \\
    Sonnet 4 & 90.70 & 82.72 & 7.97 & 93.69 & 82.06 & 11.63 & 95.02 & 77.74 & 17.28 & 76.74 & 66.11 & 10.63 & 96.01 & \textbf{86.38} & 9.63 \\
    Haiku 4.5 & 89.04 & 78.07 & 10.96 & 87.38 & 76.08 & 11.30 & 87.04 & 71.10 & 15.95 & 91.69 & 79.73 & 11.96 & 94.68 & \textbf{83.06} & 11.63 \\
    GLM-5.2 & 89.04 & 79.73 & 9.30 & 95.68 & 84.39 & 11.30 & 99.00 & 82.72 & 16.28 & 55.81 & 50.83 & 4.98 & 98.67 & \textbf{87.71} & 10.96 \\
    Qwen3-Coder-Next & 69.10 & 58.47 & 10.63 & 77.08 & 63.79 & 13.29 & 76.41 & 62.46 & 13.95 & 70.43 & 62.46 & 7.97 & 86.05 & \textbf{73.75} & 12.29 \\
    MiniMax-M2.5 & 80.40 & 71.10 & 9.30 & 82.06 & 71.10 & 10.96 & 86.05 & 71.10 & 14.95 & 52.16 & 44.85 & 7.31 & 96.35 & \textbf{83.39} & 12.96 \\
    Qwen3.6-27B & 80.07 & 69.77 & 10.30 & 83.39 & 72.43 & 10.96 & 92.03 & 75.75 & 16.28 & 61.79 & 54.82 & 6.98 & 94.68 & \textbf{82.72} & 11.96 \\
    \midrule
    \textbf{All models} & \textbf{84.93} & \textbf{74.63} & \textbf{10.30} & \textbf{88.21} & \textbf{76.33} & \textbf{11.88} & \textbf{90.53} & \textbf{74.42} & \textbf{16.11} & \textbf{72.30} & \textbf{62.92} & \textbf{9.39} & \textbf{94.89} & \textbf{83.31} & \textbf{11.59} \\
    \bottomrule
  \end{tabular}
  }%
  \endgroup
  \par\smallskip
  \begin{minipage}{\textwidth}\footnotesize
  Comp. and Exec. are compilation and execution pass rates (\%); Gap = Comp.\ $-$ Exec.\ (pp),
  computed before rounding. Each method uses 301 tasks per model
  (2,408 task--model pairs in the final row). Execution requires a complete
  scan-by-scan runtime match; all unsuccessful outcomes remain in the denominator.
  The gap includes every post-compilation failure stage as well as completed
  executions with behavioral failures. External-baseline names denote the
  public-material adaptations in Table~\ref{tab:baseline-adaptations}.
  \end{minipage}
\end{table}

Baseline outcomes comprise 1,542 without a successful compilation result
(including delivery, interface, translation, and native-build failures),
1,148 post-compilation execution failures, and 6,942 passes. The frozen
binary results do not separate all post-compilation causes across eight models.
Compilation can also change method rankings: LLM4PLC has the highest
pooled baseline compilation rate (90.53\%), but Agents4PLC has the highest
execution pass rate (76.33\%). Agents4PLC's paired gain over Direct is 1.70
points (95\% task-cluster bootstrap CI $[-0.33,3.65]$).

CycleGen uses the program returned by Direct as the first candidate in its pool. Its end-to-end
comparisons include inference-budget differences; RQ3 controls the pools
to examine component effects.

\par\smallskip
\noindent\textbf{\underline{Comparison with baselines.}}
Table~\ref{tab:study-a} reports execution pass rate under the
common final evaluator. CycleGen reaches 83.31\%, exceeding Direct by
8.68 points, adapted Agents4PLC by 6.98 points, adapted LLM4PLC by 8.89 points,
and adapted AutoPLC by 20.39 points. The 95\% task-cluster bootstrap intervals are
$[7.23,10.22]$ points against Direct and $[5.02,9.01]$ against Agents4PLC,
the strongest pooled baseline. CycleGen exceeds the best baseline for
each model by 1.66--12.29 points. Per-model differences are descriptive;
budgets differ across methods and statistical inference is reported for pooled paired contrasts.
CycleGen compiles in 2,285 of 2,408 outcomes (94.89\%), leaving an
11.59-point compile--execution gap. Thus, higher execution pass rates
do not eliminate the need for execution-based evaluation.

\par\smallskip
\noindent\textbf{\underline{SemaPLC reconstruction and single-model scope.}}
On the same 301 tasks with GPT-4o, our SemaPLC reconstruction reaches
31.89\%, 45.85\%, and 45.85\% execution pass rate at $r=0,1,2$
repairs per native check. At $r=2$, it uses 38.07 million recorded tokens,
12.82 times CycleGen's 2.97 million on that model under their respective
budgets. Native delivery accepts 75 targets; six fail the common evaluator.
Of its 138 passes, 69 are native-accepted, 42 have same-hash native
rejections, 14 are unchecked (no full native verification record), and 13
are stale (no fresh candidate/envelope verdict). The second repair adds 6.80\% tokens and rescues three outcomes
while harming three. Scaling this reconstruction to all eight models would require a costly
additional campaign, while the observed budget and delivery-gate differences
would still prevent it from forming a directly comparable method condition.
We therefore report the GPT-4o reconstruction as single-model process
evidence, rather than as an eight-model baseline ranking or a complete
reproduction claim.
\par\smallskip
\noindent\textbf{\underline{Source and scenario-length sensitivity.}}
Excluding all 126 tasks originating from Agents4PLC or AutoPLC leaves
175 tasks: the four-baseline compile--execution gap remains 11.04 points,
and CycleGen improves on Direct by 8.86 points. Separately, excluding
the six tasks with more than 1,000 scans leaves a gap of 11.33 points
and a CycleGen gain of 8.56 points. Both findings therefore persist
beyond these source groups and unusually long scenarios.

\rqanswer{1}{%
Across four baselines and eight models, execution trails compilation by
11.92 points. CycleGen raises execution pass rate from 74.63\% to 83.31\%,
but retains an 11.59-point gap: generation improvements do not replace
execution-based evaluation.
}

\subsection{RQ2: Which Failures Occur During Compilation and Execution?}
\label{sec:rq2}

For GPT-5.5, GPT-4o, Claude Sonnet~4, Claude Haiku~4.5, and GLM-5.2,
the diagnostic subset comprises 6,020 baseline outcomes. Here, delivery failures
account for 2.26\%, interface violations for 0.25\%, compilation and build
failures for 8.65\%, deployment or runtime failures after compilation for
0.03\%, and completed executions with behavioral failures for 12.39\%.
The remaining 76.41\% pass. For Direct, compilation failures and behavioral
failures account for similar proportions of all 1,505 outcomes: 9.70\% and
10.43\%, respectively; delivery failures account for 0.33\%.

Table~\ref{tab:compile-taxonomy}(a) groups all Direct compilation failures
by their first diagnostic, avoiding repeated counts of cascading messages.
These are reported symptoms, not definitive root causes~\cite{yasunaga2020drrepair}.

\begin{table}[!htbp]
  \centering
  \caption{Direct failure categories among compiler errors (a, $n=146$)
  and completed executions with mismatched outputs (b, $n=157$).}
  \label{tab:compile-taxonomy}
  \begingroup
  \fontsize{8.5}{10.2}\selectfont
  \renewcommand{\arraystretch}{1.12}
  \setlength{\tabcolsep}{2pt}
  \begin{minipage}[t]{0.49\linewidth}
    \centering
    \textbf{(a) First compiler diagnostic}\par\smallskip
    \begin{tabular}{@{}p{0.68\linewidth}r@{}}
      \toprule
      Category & $n$ (\%) \\
      \midrule
      Malformed statements & 41 (28.08) \\
      Malformed declarations & 37 (25.34) \\
      Type incompatibility & 21 (14.38) \\
      Malformed expressions & 21 (14.38) \\
      Invalid variable references & 21 (14.38) \\
      Block-interface misuse & 4 (2.74) \\
      Unresolved syntax diagnostic & 1 (0.68) \\
      \bottomrule
    \end{tabular}
  \end{minipage}\hfill
  \begin{minipage}[t]{0.49\linewidth}
    \centering
    \textbf{(b) Reviewed behavior mismatch}\par\smallskip
    \begin{tabular}{@{}p{0.68\linewidth}r@{}}
      \toprule
      Category & $n$ (\%) \\
      \midrule
      Public-output updates & 43 (27.4) \\
      Arithmetic / boundary logic & 35 (22.3) \\
      Timing / call order & 25 (15.9) \\
      Reset / priority & 15 (9.6) \\
      State retention / overwrite & 14 (8.9) \\
      Edge detection & 13 (8.3) \\
      Initialization & 9 (5.7) \\
      Numerical precision / boundary & 3 (1.9) \\
      \bottomrule
    \end{tabular}
  \end{minipage}
  \endgroup
  \par\smallskip
  \begin{minipage}{\linewidth}\footnotesize
  Panel (a) assigns the first compiler diagnostic to each of 146
  failures; these are symptoms, not established root causes.
  Panel (b) gives one reviewer-assigned primary category to each of 157
  behavioral failures. Its three numerical cases retain the mismatch
  verdict but are separate from the 154 implementation-error cases.
  \end{minipage}
\end{table}

Malformed statements and declarations account for 53.42\% of compiler
failures; type incompatibility accounts for 14.38\%.
The remaining categories cover expressions, references, block interfaces,
and one unresolved diagnostic. These symptoms alone cannot establish
behavioral correctness.

Table~\ref{tab:compile-taxonomy}(b) covers all 157 Direct behavioral
failures: 154 reviewer-attributed implementation errors and three
numerical cases. Public-output updates (43, 27.4\%), arithmetic or
boundary logic (35, 22.3\%), and timing or call order (25, 15.9\%)
together account for 65.6\%. Numerical cases retain their mismatch
verdicts separately from implementation mechanisms. RQ4's
Fig.~\ref{fig:rq4-mechanisms} relates all categories to observation
policies.

The independent annotation check agrees on 33 of 40 outcomes (82.5\%),
with unweighted Cohen's $\kappa=0.78$ (95\% bootstrap CI $[0.61,0.91]$).
Public-output updates are the largest category in both labelings. Six of
the seven disagreements concern this broad category versus a more specific
mechanism, illustrating the difficulty of assigning one primary label to
programs that may contain several defects.

Timing/call-order failures include invoking a stateful block before its
current-scan inputs are assigned, so it consumes previous input values.
RQ4 presents a traced example (Fig.~\ref{fig:behavior-witnesses}).

\rqanswer{2}{%
Statements and declarations account for 53.42\% of Direct compilation
failures; output-update, arithmetic, and timing/call-order errors account
for 65.6\% of behavioral failures. Runtime checks reveal state and
value errors that compiler feedback cannot establish.
}

\subsection{RQ3: How Do CycleGen's Modules Contribute to Execution Correctness?}
\label{sec:rq3}

RQ1 measures overall effectiveness; here we examine which parts of
CycleGen produce its gains and at what token cost. Table~\ref{tab:framework-pairs}
ablates pool repair and self-test selection on the same five initial
candidates. Removing Compile-Fix retains the shared suite and selector;
removing self-test selection retains the repaired pool but returns the
first eligible candidate. Self-Test and Selection Gate are evaluated
jointly, since runtime observations guide the gate's ranking; their
individual effects are not separately identified. We call these variants
Self-test only and Repair only, respectively: ``only'' refers to the two
optional components added to the multiple-candidate configuration, and
Self-test only includes execution-based selection.

\begin{table}[!htbp]
  \centering
  \caption{CycleGen ablations on 2,408 task--model pairs.}
  \label{tab:framework-pairs}
  \small
  \setlength{\tabcolsep}{3pt}
  \renewcommand{\arraystretch}{1.14}
  \begin{tabular*}{\linewidth}{@{\extracolsep{\fill}}lcllrrr@{}}
    \toprule
    Configuration & $k$ & Repair & Selection & Comp. & Exec. & Tokens \\
    \midrule
    Direct & 1 & No & First & 84.93 & 74.63 & 1,929 \\
    Multiple candidates & 5 & No & First & 92.23 & 79.94 & 9,740 \\
    Self-test only & 5 & No & Self-test & 92.23 & 81.02 & 16,015 \\
    Repair only & 5 & Yes & First & 94.85 & 81.69 & 11,922 \\
    CycleGen & 5 & Yes & Self-test & 94.89 & 83.31 & 18,197 \\
    \bottomrule
  \end{tabular*}
  \par\smallskip
  \begin{minipage}{\linewidth}\footnotesize
  Comp. measures complete builds in the common evaluator; Exec. measures
  execution passes (\%). Tokens is mean recorded
  input plus output tokens per task--model pair, rounded to integers.
  First selects the first candidate preserving the interface and passing
  ST-to-C translation; Direct returns the first
  generated candidate. All five-candidate configurations reuse the same
  initial samples, and self-test configurations reuse the same suite.
  Counts include unsuccessful outcomes.
  \end{minipage}
\end{table}

\par\smallskip
\noindent\textbf{\underline{Context \& Generation:}} Supplying candidate alternatives.
First-eligible selection from five unrepaired candidates reaches 79.94\%,
a 5.32-point gain over Direct, while mean tokens increase from 1,929 to
9,740. This contrast changes sampling budget, not context or prompt.
The remaining modules rescue 95 outcomes and break 14, adding 3.36
points on these pools (95\% task-cluster bootstrap CI $[2.37,4.40]$).

\par\smallskip
\noindent\textbf{\underline{Compile-Fix:}} Repairing front-end failures.
Removing compiler-guided repair~\cite{yasunaga2020drrepair} while fixing
the suite and selector lowers execution pass rate by 2.28 points
(95\% CI $[1.45,3.16]$) and compilation from 94.89\% to 92.23\%.
Repair admits rejected candidates to runtime testing without establishing
correct behavior; its mean-token increment is 2,183.

\par\smallskip
\noindent\textbf{\underline{Self-Test and Selection Gate:}} Selecting by execution feedback.
Self-test selection~\cite{chen2022codet,li2024doce} rescues 49 outcomes
and breaks ten versus first-eligible selection on repaired pools:
a 1.62-point gain (95\% CI $[0.91,2.37]$; Holm-adjusted
$p=1.14\times10^{-5}$), for mean tokens rising from 11,922 to 18,197.
Its one-outcome compilation gain occurs when the first eligible candidate
passes ST-to-C translation but fails native building; completed self-tests
favor another candidate. The contrast therefore includes build validation
as well as behavioral ranking. On unrepaired pools, selection adds 1.08
points ($[0.50,1.70]$).

\par
\begin{figure}[!htbp]
  \centering
  \includegraphics[width=\linewidth]{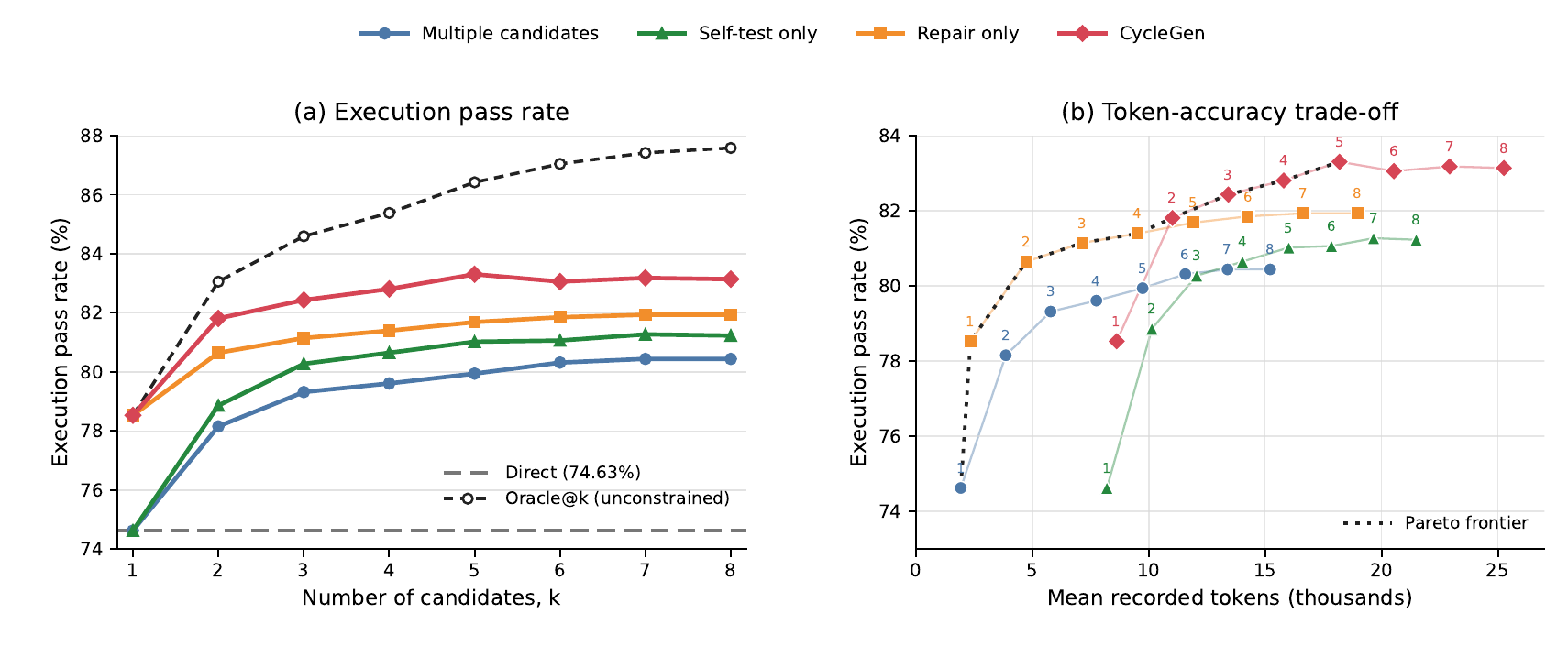}
  \caption{Ordered-prefix component sensitivity (301 tasks $\times$ eight
  models). The left panel shows the four controlled configurations,
  Direct, and the unconstrained repaired-pool Oracle@$k$. Self-test only
  retains the shared suite and selector without pool repair; Repair only
  repairs the pool and selects its first eligible candidate. Both use
  multiple candidates. The right panel plots
  pass rate against mean input-plus-output tokens; point labels give
  $k$, and the dotted curve marks the observed Pareto frontier. For
  Oracle@$k$, percentages count hidden behavioral matches among frozen pool
  candidates without public-interface eligibility, not formal task passes;
  this unconstrained diagnostic is not a deployable selector.}
  \label{fig:component-k}
  \Description{The left panel plots four execution-pass-rate curves from
  one to eight candidates, a Direct horizontal reference, and a varying
  unconstrained Oracle-at-k diagnostic ceiling. The right panel plots execution pass rate against
  mean recorded tokens for all method and candidate-count configurations,
  with candidate count labels and a dotted Pareto frontier.}
\end{figure}

\smallskip
\noindent\textbf{\underline{Candidate count and token trade-offs.}}
Fig.~\ref{fig:component-k} follows frozen generation-order prefixes
$k=1,\ldots,8$ and their token costs. Repaired $k=1$ already includes
Compile-Fix and differs from Direct. CycleGen rises from 78.53\% at
$k=1$ to 83.31\% at $k=5$. From $k=3$ to $5$, it gains 21 passes
(0.87 points) for 35.6\% more tokens (13,419.9 to 18,197.2).
Extending to $k=8$ raises tokens to 25,256.6 (+38.8\%) but loses four
selected passes (0.17 points). Observed $k=5$ dominates its $k=6$--8
extensions; this does not establish a generally optimal or equal-resource
budget.

At $k=5$, CycleGen averages 7 model calls and 18,197 tokens per task-model
pair; Direct averages 1 call and 1,929 tokens. CycleGen’s 16,864 calls comprise 12,040 candidate generations, 3,124 test-authoring calls, and 1,700 repairs.
Counts include shared Direct dependencies and each suite once per pool;
local execution, dollar costs, and latency are outside this measure.

\par\smallskip
\noindent\textbf{\underline{Candidate availability and selection limits.}}
Oracle@$k$ retrospectively counts hidden behavioral matches without
public-interface eligibility; these diagnostic labels neither guide
selection nor denote formal task passes. The unconstrained repaired-pool
oracle in Fig.~\ref{fig:component-k}(a)
rises from 2,081/2,408 (86.42\%) at $k=5$ to 87.58\% at $k=8$, although
CycleGen's selected rate does not improve. At $k=5$, 75 pools contain unselected recorded behavioral matches, while 327 lack any documented match.
Six unconstrained-oracle witnesses violated the public interface contract
and were therefore ineligible for deployment; we retain them only to
distinguish feasibility losses from selection-stage misses.
The remaining 69 involve the selection stage: 15 score ties, 21 cases
favoring a failing candidate with more passing self-test cases, seven
favoring one with fewer output mismatches, 16 with invalid self-test
execution, and ten without a scored suite. Thus, the oracle gap includes
both interface-contract violations and limitations of self-test coverage,
execution availability, and ranking. In 13 of 15 tied pools, the passing
witness and selected failure have identical public binding observations;
changing expected outputs alone cannot separate them. This motivates the
following diagnostic of self-test agreement and discrimination, which is
supporting process evidence rather than an additional module ablation.
\par\smallskip
\noindent\textbf{\underline{Self-test agreement and discrimination.}}
A separate 24-task pilot examines the quality of the self-test signal.
Tasks were selected by source rotation and reference-length/static-feature
strata, independently of candidate success and self-test outcomes.
The pilot covers 192 task--model suite slots: 144 pass the suite-validity
checks, 46 are invalid, and two are missing or unparseable. Of the 144
valid suites, 138 complete repeatable reference execution. The other six
reach a specified nonterminating input on one task; that task's hidden
evaluation scenario exercises only terminating inputs.
Among the 138 repeatable suites, original
assertions match 2,242/2,673 binding scans (83.88\%) but accept all binding
cases in only 75/138 suites (54.35\%). We apply 50 single-site changes to comparison boundaries, equality,
logical connectives, and Boolean literals across 18 tasks, with at most
three mutants per task. A detection requires the same suite to accept
the reference and reject the mutant after complete stable execution.
Among 135 stable suite--mutant pairs
whose reference is accepted, 99 reject the mutant (73.33\%; 95\% CI
$[51.04,91.30]$); only 43/192 original suite slots detect any mutant.

\par\smallskip
\noindent\textbf{\underline{Scope of the ablations.}}
These exploratory comparisons fix pools and stated controls, not total
inference resources. 

\rqanswer{3}{%
Multiple candidates raise execution pass rate from 74.63\% to 79.94\%
with front-end-based selection. On controlled pools, Compile-Fix adds
2.28 points with self-test selection fixed, while Self-Test and Selection
Gate add 1.62 points with repair fixed. These conditional gains show the
roles of candidate diversity, front-end repair, and execution-based selection.
Increasing $k$ from five to eight adds 38.8\% tokens without improving
selected pass rate, despite a higher unconstrained pool-oracle rate.
}

\subsection{RQ4: How Much Does Observation Granularity Change Correctness Verdicts?}
\label{sec:rq4}

We subsample Direct traces with programs, inputs, output sets, and decoded
type-and-value equality fixed. Unlike expanding test inputs~\cite{liu2023evalplus},
this asks when outputs must be observed within the same execution.

Validated replays cover all 157 Direct behavioral failures detected by
per-scan comparison. Here, a missed failure means a disagreement detected
by that criterion but not by the coarser policy. Exact passes also pass
every subset of the same observations.

Checking every second, fifth, or tenth scan misses 11, 34, or 47 of
157 behavioral failures (7.01\%, 21.66\%, and 29.94\%), respectively,
even when each scenario's first and final scans are always checked.
The apparent execution pass rate across 1,505 Direct outcomes rises from
79.53\% with every-scan checking to 80.27\%, 81.79\%, and 82.66\%.
Final-scan-only observation misses 98 failures (62.42\%; 95\%
task-cluster CI 50.00--74.14\%) and raises the apparent pass rate to
86.05\%. These comparisons use the same complete traces; only the
observation positions change.

\par\smallskip
\noindent\textbf{\underline{Transient disagreements hidden by later outputs.}}
The shift register in Fig.~\ref{fig:motivation} illustrates how reset
conceals incorrect intermediate values. Fig.~\ref{fig:behavior-witnesses}
shows the same observation problem for a timer: the candidate invokes the
off-delay block before assigning its current sensor input. The light is
wrong at scan~7 but matches again at scans~8--9. Observing only these later
matching outputs misses the earlier disagreement; comparing each scan
exposes the one-scan delay.

\begin{figure}[!htbp]
  \centering
  \begin{minipage}[c]{0.57\linewidth}
  \begin{lstlisting}[basicstyle=\ttfamily\scriptsize,xleftmargin=0pt,frame=single,breaklines=true]
(* Candidate: uses previous input *)
TOF0(IN := pir_active, PT := T#20s);
(* ... manual-control updates ... *)
pir_active := stairs_pir_sensor
              AND NOT lights_buttons_state;
stairs_light := lights_buttons_state
                OR TOF0.Q;
  \end{lstlisting}
  \end{minipage}\hfill
  \begin{minipage}[c]{0.40\linewidth}
  \centering\scriptsize
  \begin{tabular}{@{}lcccc@{}}
  \toprule
  Scan & 6 & 7 & 8 & 9 \\
  \midrule
  Motion input & 0 & 1 & 1 & 0 \\
  Reference light & 0 & \textbf{1} & 1 & 1 \\
  Candidate light & 0 & \textbf{0} & 1 & 1 \\
  Output match & Yes & No & Yes & Yes \\
  \bottomrule
  \end{tabular}
  \end{minipage}
  \caption{A transient timer disagreement missed by later observations.
  The candidate calls its off-delay timer before updating the current input.
  With manual mode off, the recorded Boolean trace (0/1) differs at scan~7
  and agrees at scans~8--9. Code is excerpted; bold marks the disagreement.}
  \Description{Candidate code invokes the timer before updating its input.
  Reference and candidate lights disagree at scan seven and agree again
  at scan eight, concealing the transient delay in later observations.}
  \label{fig:behavior-witnesses}
\end{figure}

\par
\begin{figure}[tb]
  \centering
  \includegraphics[width=0.95\linewidth]{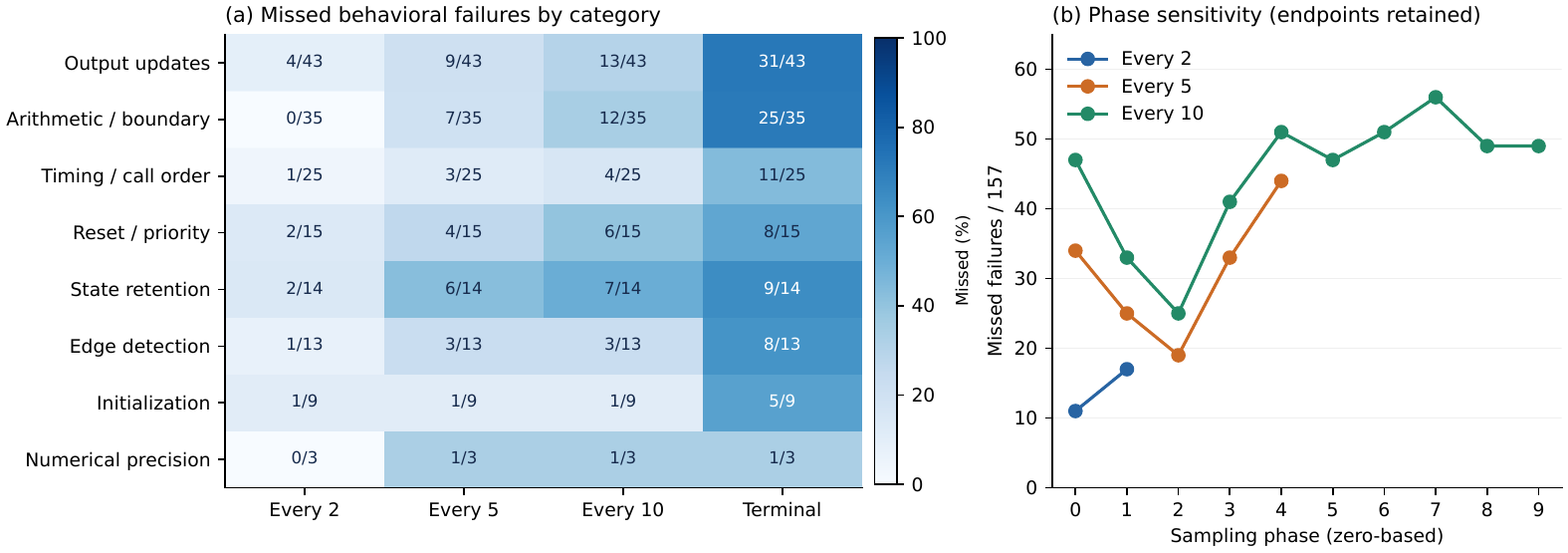}
  \caption{Observation-granularity mechanisms on 157 five-model Direct
  behavioral failures. Left: missed/total within each audited category;
  color denotes percentage. Right: all fixed sampling phases, retaining
  each scenario's endpoints. This post-hoc analysis keeps program traces fixed.}
  \label{fig:rq4-mechanisms}
  \Description{A matrix gives missed failures by error category and
  observation policy. A phase plot shows variation in missed failures
  for strides two, five, and ten; no fixed phase detects all failures.}
\end{figure}

\smallskip
\noindent\textbf{\underline{Mechanisms and within-scenario re-agreement.}}
Joining the 157 failures to audited labels by task, model, and candidate
hash shows that final-scan-only checking misses 9/14 state-retention,
8/13 edge-detection, and 31/43 output-update failures
(Fig.~\ref{fig:rq4-mechanisms}). Overall, 144/157 failures exhibit
within-scenario output re-agreement (95\% task-cluster interval:
85.31--96.84\%), including all 98 final-scan misses; in 115/157 failures, no mismatch run exceeds five consecutive scans, though some final-scan misses have longer runs.
Cross-scenario resets do not count as re-agreement; visible agreement
does not establish internal-state equivalence.

\par\smallskip
\noindent\textbf{\underline{Sensitivity to sampling phase.}}
For each stride $s\in\{2,5,10\}$, we apply every possible phase uniformly
across scenarios, retaining their endpoints. Phase zero yields 11, 34, and
47 missed failures; across phases, these counts range over 11–17, 19–44, and 25–56, respectively.
Equally weighting the $s$ phases gives mean missed counts of
14.0, 31.0, and 44.9 of the same 157 failures, respectively.
Every fixed phase misses failures. Endpoint retention slightly changes
observation budgets. Full-trace dominance is structural; these results
quantify its magnitude, mechanisms, and phase sensitivity on fixed traces.

\par\smallskip
\noindent\textbf{\underline{Implications for execution-based evaluation.}}
RQ4 changes observation positions, not programs or input histories.
Higher apparent pass rates therefore reflect missed evidence rather than
better code. Endpoint retention and phase changes leave failures undetected
on these traces. Per-scan observation preserves exposed disagreements;
broader input scenarios address a separate coverage question. Neither
establishes correctness on executions that were not run.

\rqanswer{4}{%
Every-second and every-fifth-scan checking miss 7.01\% and 21.66\% of
behavioral failures. Transient discrepancies can disappear before an
observation; sampling phase changes which failures are missed.
Per-scan comparison detects them on the supplied executions.
}

\section{Threats to Validity}
\label{sec:threats}

\noindent\textbf{\underline{Construct validity.}}
Execution pass rate measures reference agreement on supplied scenarios,
not full requirement coverage, unseen inputs, physical behavior, or
termination. Exact equality may count numerical rounding differences as failures. Two PLC engineers
checked requirements against sources, interfaces, and available reference
behavior. RQ2 assigns one primary label to each behavioral failure; independent annotation
assesses consistency, not unique root causes or sparse-category accuracy
(Section~\ref{sec:rq2}).

\par\smallskip
\noindent\textbf{\underline{Baseline fidelity.}}
Agents4PLC and LLM4PLC retain compiler-feedback repair but omit described
formal-verification loops; common scoring cannot restore generation-time
verification. Results characterize these adaptations, not native systems.
Sampling and inference budgets differ; shared pools narrow but do not remove
this limitation.

\par\smallskip
\noindent\textbf{\underline{Internal validity.}}
Reference self-checks, fresh state, aligned logical time, hidden-data
isolation, and versioned images constrain initialization, timing, and
evaluator effects.

\par\smallskip
\noindent\textbf{\underline{External validity.}}
Eight deployments cover 301 tasks; taxonomy and observation analyses
cover five models. Source concentration and limited scenarios constrain
diversity and coverage~\cite{allamanis2019duplication}
(Table~\ref{tab:benchmark-characteristics}). Baseline-derived tasks and
Qwen3-Coder-Next's roles in scenario authoring and evaluation may affect
results; training exposure remains unknown. Fixed budgets, one runtime,
and provider-specific reasoning limit transfer to other dialects and
physical controllers (Section~\ref{sec:design}). AutoPLC's lower compilation
under our adapter cannot be attributed to individual components.

\par\smallskip
\noindent\textbf{\underline{Conclusion validity.}}
Task-level paired resampling preserves method--model correlation. The CycleGen configuration was developed using this benchmark. Independent tasks and repeated generations are needed to assess generalization and sampling stability. Ablations estimate
conditional effects under the reported controls.
\section{Conclusion}
\label{sec:conclusion}

CycleBench evaluates LLM-generated ST through scan-by-scan execution on 301 PLC tasks. Across eight models and four baseline methods, execution trails compilation by 11.92 percentage points. Every-fifth-scan checking misses 21.66\% of 157 behavioral failures in the five-model Direct analysis, even with scenario endpoints retained; transient discrepancies and sampling phase explain these misses. CycleGen combines multiple candidates, compiler-guided repair, and runtime self-tests, raising execution pass rate from 74.63\% to 83.31\%. Controlled-pool ablations support repair and self-test selection, while increasing the candidate count beyond five adds token cost without improving execution pass rate. These findings motivate scan-by-scan evaluation and behavioral feedback for generated PLC programs.

\label{sec:conclusion-end}
\clearpage \section*{Data Availability} We plan to publicly release the code and data after the peer-review process is complete.
\label{sec:body-end}


\bibliographystyle{ACM-Reference-Format}
\bibliography{refs}

\end{document}